\documentclass[prl,twocolumn,showpacs,preprintnumbers,amsmath,amssymb,floats]{revtex4-1}
\usepackage{graphicx}
\usepackage{amsmath}
\usepackage{epstopdf}
\usepackage{amsbsy}
\usepackage{color}
\usepackage{dcolumn}
\usepackage{bm}

\begin{document}

\title{Enhancing magnetic stripe order in iron pnictides by RKKY exchange interactions}
\author{Maria N. Gastiasoro and Brian M. Andersen}
\affiliation{Niels Bohr Institute, University of Copenhagen, Universitetsparken 5, DK-2100 Copenhagen,
Denmark}

\date{\today}

\begin{abstract}

Recent experimental studies have revealed several unexpected properties of Mn-doped BaFe$_2$As$_2$. These include extension of the stripe-like magnetic $(\pi,0)$ phase to high temperatures above a critical Mn concentration only, the presence of diffusive and weakly temperature dependent magnetic $(\pi,\pi)$ checkerboard scattering, and an apparent absent structural distortion from tetragonal to orthorhombic. Here, we study the effects of magnetic impurities both below and above the N\'{e}el transition temperature within a real-space five-band model appropriate to the iron pnictides. We show how these experimental findings can be explained by a cooperative behavior of the magnetic impurities and the conduction electrons mediating the Ruderman-Kittel-Kasuya-Yosida (RKKY) interactions between them.

\end{abstract}


\maketitle

Whether the electronic fluctuations in Fe-based superconductors are predominantly of magnetic or orbital nature remains controversial.\cite{fernandes14} The structural transition from tetragonal to orthorhombic symmetry has been argued to arise from an electronically driven nematic instability caused either by orbital order\cite{kruger09,chen09,lv09,lee09} or so-called spin-Ising order.\cite{fang08,xu08,hu12,fernandes12} These order parameters are, however, intimately linked by symmetry and cannot exist on their own,\cite{fernandes14} making it hard to determine experimentally which order exhibits the dominant susceptibility in the high-$T$ normal phase and hence drive e.g. the structural transition. A resolution to this question is of great interest since the dominant fluctuations are likely to also mediate the pairing required for superconductivity. Therefore, the presence of a magnetic tetragonal phase has attracted a lot of attention\cite{kim10,avci14}; this phase exhibits magnetic order at the same ordering vectors $(\pi,0),(0,\pi)\equiv\mathbf{Q}_{stripe}$ as the standard stripe magnetic order, but without a concomitant breaking the tetragonal symmetry. The existence of magnetic non-orthorhombic phases argues against orbital order as the driving instability of these materials, and has been interpreted in terms of $C_4$ symmetric magnetic structures with ordering at  $\mathbf{Q}_{stripe}$.\cite{lorenzana08,eremin2010} 

For the pnictides, the above issues have largely focussed on hole-doped BaFe$_2$As$_2$ where particularly Ba(Fe$_{1-x}$Mn$_x$)$_2$As$_2$ constitutes an interesting case in point. For this compound, Kim {\it et al.}\cite{kim10} found that the structural phase transition disappears at a critical Mn doping ($x_c\sim0.1$) whereas the $\mathbf{Q}_{stripe}$ magnetic order remains. Remarkably, at higher doping than $x_c$ the magnetic order exhibits a new high-temperature component as seen by the persistence of a broad $\mathbf{Q}_{stripe}$ magnetic Bragg peak well in excess of the N\'{e}el temperature $T_N$ of the lower doped $x<x_c$ samples. In addition, inelastic neutron scattering revealed that Mn dopants induce short-range quasi-elastic spin scattering at $(\pi,\pi)\equiv \mathbf{Q}_{N\acute{e}el}$, which persists at all measured $T$, and coexists with the long-range ordered $\mathbf{Q}_{stripe}$ phase at low $T$.\cite{tucker12} The presence of local antiferromagnetic (AF) correlations induced by Mn ions have also been detected by nuclear magnetic resonance (NMR).\cite{texier12,leboeuf13} These studies therefore suggest that Mn locally nucleate magnetic moments consistent with $\mathbf{Q}_{N\acute{e}el}$ structure. 

More recently, Inosov {\it et al.}\cite{inosov13} performed a comprehensive mapping of the phase diagram of Ba(Fe$_{1-x}$Mn$_x$)$_2$As$_2$  in the range $0<x<0.12$. The existence of the novel high-$T$ stripe magnetic phase at $x>x_c$ was confirmed by these studies. However, by combining neutron data with muon spin relaxation ($\mu$SR) and NMR measurements, Inosov {\it et al.}\cite{inosov13} proposed a spatially inhomogeneous picture where Mn ions act as magnetic impurities that induce $(\pi,0)$ magnetic rare regions already above $T_N$ of the parent compound, and with a volume fraction of these rare regions growing continuously with decreasing $T$. In contrast to the earlier study of Ref.~\onlinecite{kim10}, a finite orthorhombic distortion, most likely associated with the stripe-like magnetic rare regions, was shown to coexist with regions of tetragonal lattice symmetry.\cite{inosov13}

To help resolve the controversy of Ba(Fe$_{1-x}$Mn$_x$)$_2$As$_2$, and understand, more generally, the physics of magnetic disorder in iron pnictides, microscopic theoretical model calculations are highly called for. Minimal requirements of such a theoretical description include being able to 1) explain how magnetic impurities can generate long-range magnetic $\mathbf{Q}_{stripe}$ order at high $T>T_N$, 2) explain why this happens only above a certain critical concentration of magnetic disorder, 3) explain the presence of diffusive $\mathbf{Q}_{N\acute{e}el}$ magnetic scattering at low concentrations, and 4) explain why the orthorhombicity appears absent at high enough impurity concentrations. In addition, the spatial modulations evidenced by the data of Ref.~\onlinecite{inosov13} points to the importance of a real-space approach and a careful study to cooperative impurity effects in these systems.

Here, we provide such a microscopic real-space description of magnetic disorder relevant to iron pnictides. We use a realistic five-band model with standard onsite Coulomb repulsion to study the induced magnetic order nucleated by magnetic impurity sites. It is found that magnetic impurities exhibit a $\mathbf{Q}_{N\acute{e}el}$ magnetic structure close to its core, as well as longer-ranged magnetic tails of $\mathbf{Q}_{stripe}$ modulations which may overlap with neighbouring impurities and induce long-range $\mathbf{Q}_{stripe}$ magnetic order even above $T_N$ of the clean system. This cooperative effect only takes place, however, if the length scale of the magnetic impurity tails is comparable to the average inter-impurity distance, yielding a natural explanation for the detected critical concentration $x_c$. In fact, as will be shown in detail below, all four criteria above are contained within our model. At higher Mn concentrations we predict that the induced $\mathbf{Q}_{stripe}$ order vanishes because there is no room on average to host this order, and only the  $\mathbf{Q}_{N\acute{e}el}$ magnetic structure remains. This crossover happens well before reaching the clean system BaMn$_2$As$_2$ which is known to exhibit $\mathbf{Q}_{N\acute{e}el}$ order in its ground state.\cite{singh09}

The study presented here provides an alternative scenario for the magnetism of Ba(Fe$_{1-x}$Mn$_x$)$_2$As$_2$ as compared to previous Landau models assuming homogeneous phases.\cite{lorenzana08,eremin2010,wang13} In addition, previous calculations of impurity-induced magnetism in Fe-based superconductors have focussed on the role of potential scatterers,\cite{gastiasoro13,gastiasoroprb13,gastiasoroprb14}, but none have studied the collective effects of magnetic impurities within a microscopic approach.
 
Finally, we note that this problem constitutes an interesting example of the more general problem of RKKY exchange interactions in multi-orbital nested systems at the brink of an instability,\cite{aristov97,akbari11} and the physics of AF rare regions in itinerant systems.\cite{vojta10} In the standard case of magnetic impurities in metallic hosts, the conduction elections are usually integrated out, giving rise only to the RKKY effective interaction between the impurity spins which may lead to interesting RKKY spin-glass behavior.\cite{binder86,fischer99} Here, however, we do not integrate out the itinerant electrons since their response to the magnetic impurities is crucial for explaining the measurements discussed above. 

The five-orbital Hamiltonian is given by
\begin{equation}
 \label{eq:H}
 \mathcal{H}=\mathcal{H}_{0}+\mathcal{H}_{int}+\mathcal{H}_{imp},
\end{equation}
with $\mathcal{H}_0$ the kinetic part 
\begin{equation}
 \label{eq:H0}
\mathcal{H}_{0}=\sum_{\mathbf{ij},\mu\nu,\sigma}t_{\mathbf{ij}}^{\mu\nu}c_{\mathbf{i}\mu\sigma}^{\dagger}c_{\mathbf{j}\nu\sigma}-\mu_0\sum_{\mathbf{i}\mu\sigma}n_{\mathbf{i}\mu.\sigma}.
\end{equation}
The operators $c_{\mathbf{i}\mu\sigma}$ ($c_{\mathbf{i} \mu\sigma}^{\dagger}$) annihilate (create) an electron at site $i$ with spin $\sigma$ in orbital state $\mu$, and $\mu_0$ is the chemical potential fixed to yield a doping level of $\delta=\langle n \rangle - 6.0 = 0$ since Mn ions do not dope the system.\cite{texier12,leboeuf13,suzuki13} 
The indices $\mu$ and $\nu$ denote  the five iron orbitals $d_{xz}$, $d_{yz}$, $d_{x^2-y^2}$, $d_{xy}$, and $d_{3z^2-r^2}$. 
The hopping amplitudes $t_{\mathbf{ij}}^{\mu\nu}$ are identical to those provided in Ref. \onlinecite{ikeda10}. 
The second term in Eq.(\ref{eq:H}) describes the onsite Coulomb interaction
\begin{align}
 \label{eq:Hint}
 \mathcal{H}_{int}&=U\sum_{\mathbf{i},\mu}n_{\mathbf{i}\mu\uparrow}n_{\mathbf{i}\mu\downarrow}+(U'-\frac{J}{2})\sum_{\mathbf{i},\mu<\nu,\sigma\sigma'}n_{\mathbf{i}\mu\sigma}n_{\mathbf{i}\nu\sigma'}\\\nonumber
&\quad-2J\sum_{\mathbf{i},\mu<\nu}{\bf S}_{\mathbf{i}\mu}\cdot {\bf S}_{\mathbf{i}\nu}+J'\sum_{\mathbf{i},\mu<\nu,\sigma}c_{\mathbf{i}\mu\sigma}^{\dagger}c_{\mathbf{i}\mu\bar{\sigma}}^{\dagger}c_{\mathbf{i}\nu\bar{\sigma}}c_{\mathbf{i}\nu\sigma},
\end{align}
which includes the intraorbital (interorbital) interaction $U$ ($U'$), the Hund's rule coupling $J$ and the pair hopping energy $J'$.
We assume spin rotation invariance $U'=U-2J$ and $J'=J$. In this work we additionally take $U=1.2$ eV and $J=U/6$, which leads to an ordered moment of the right magnitude relevant for BaFe$_2$As$_2$. Mn ions in BaFe$_2$As$_2$ are known to carry a large local magnetic moment\cite{texier12,leboeuf13,yang13} ${\bf S}$ which interacts with the spin density of the itinerant electrons
\begin{equation}\label{Himp}
 \mathcal{H}_{imp}=J_0\sum_{\{\mathbf{i^*}\}\mu\sigma\sigma'}{\bf S}_{\mathbf{i^*}} \cdot (c_{\mathbf{i^*}\mu\sigma}^{\dagger} \mbox{\boldmath{$\sigma$}}_{\sigma\sigma'} c_{\mathbf{i^*}\mu\sigma'}),
\end{equation}
where $\{\mathbf{i^*}\}$ denotes the sub-set of lattice sites containing impurity spins. In the present work, we neglect orbital dependence and spin-flip processes which reduces Eq.~(\ref{Himp}) to an Ising coupling $\sum_{\{\mathbf{i^*}\}\mu\sigma}J_0S^z_{\mathbf{i^*}\sigma} c_{\mathbf{i^*}\mu\sigma}^{\dagger}  c_{\mathbf{i^*}\mu\sigma}$. Additionally we assume the large spin (classical) limit where $J_0S\rightarrow \infty$, and Eq.~(\ref{Himp}) reduces to that of a spin dependent potential.\cite{shiba68,balatsky06}

After a mean-field decoupling of Eq.\eqref{eq:Hint}, we solve the eigenvalue problem $\sum_{\mathbf{j}\nu}
H^{\mu\nu}_{\mathbf{i} \mathbf{j} \sigma}
 u_{\mathbf{j}\nu}^{n}
=E_{n} u_{\mathbf{i}\mu}^{n}$,
where
\begin{align}
 H^{\mu\nu}_{\mathbf{i} \mathbf{j} \sigma}&=t_{\mathbf{ij}}^{\mu\nu}+\delta_{\mathbf{ij}}\delta_{\mu\nu}[-\mu_0+\delta_{\mathbf{ii^*}}J_0S_{\mathbf{i^*}\sigma}^{z}+U \langle n_{\mathbf{i}\mu\bar{\sigma}}\rangle\\\nonumber
\quad&+\sum_{\mu' \neq \mu}(U'\langle n_{\mathbf{i}\mu' \bar{\sigma}}\rangle+(U'-J)\langle n_{\mathbf{i}\mu' \sigma}\rangle)],
 \end{align}
on $N_x\times N_y$ lattices with self-consistently obtained densities $\langle n_{\mathbf{i}\mu\sigma} \rangle=\sum_{n}|u_{\mathbf{i}\mu\sigma}^{n}|^{2}f(E_{n\sigma})$, and $f(E)$ denoting the Fermi function. 

When multiple impurities are included at different randomly chosen sites, the relative signs of the individual impurity spins $S^z_{\mathbf{i^*}\sigma}$ become important and are obtained by minimising the free energy $\mathcal{F}=\mathcal{U}-T\mathcal{S}$.
Here the internal energy $\mathcal{U}=\langle\mathcal{H}^{MF}\rangle=\langle\mathcal{H}_0\rangle+\langle\mathcal{H}_{int}^{MF}\rangle+\langle\mathcal{H}_{imp}\rangle$, and the entropy $\mathcal{S}$ is obtained from the expression
\begin{equation}
 \mathcal{S}\!=\!-k_B \sum_n \left[ f(E_n)\ln f(E_n) +  f(-E_n)\ln f(-E_n)\right].
\end{equation}

\begin{figure}[t]
\begin{center}
\includegraphics[width=8.2cm]{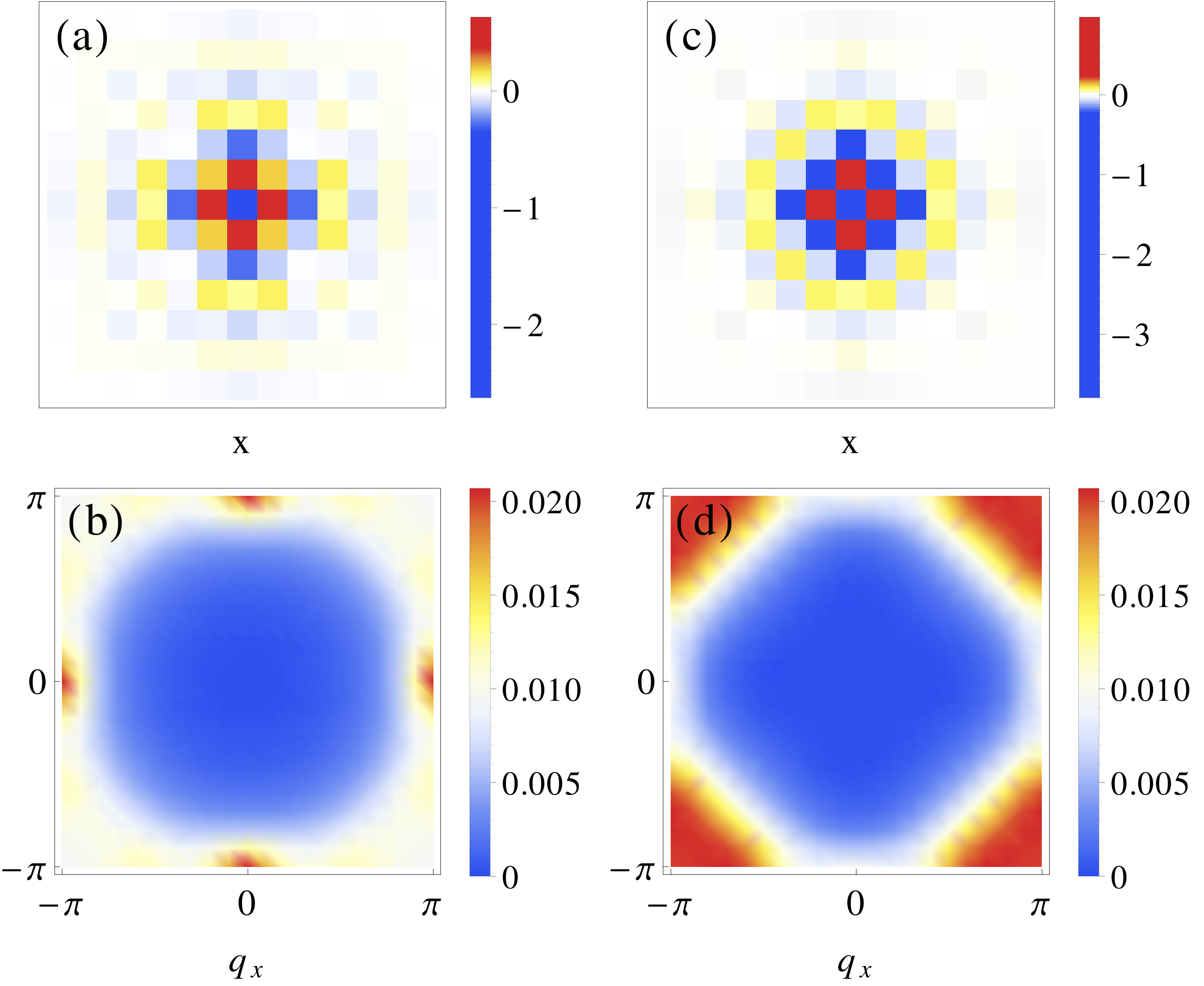}
\includegraphics[width=8.2cm]{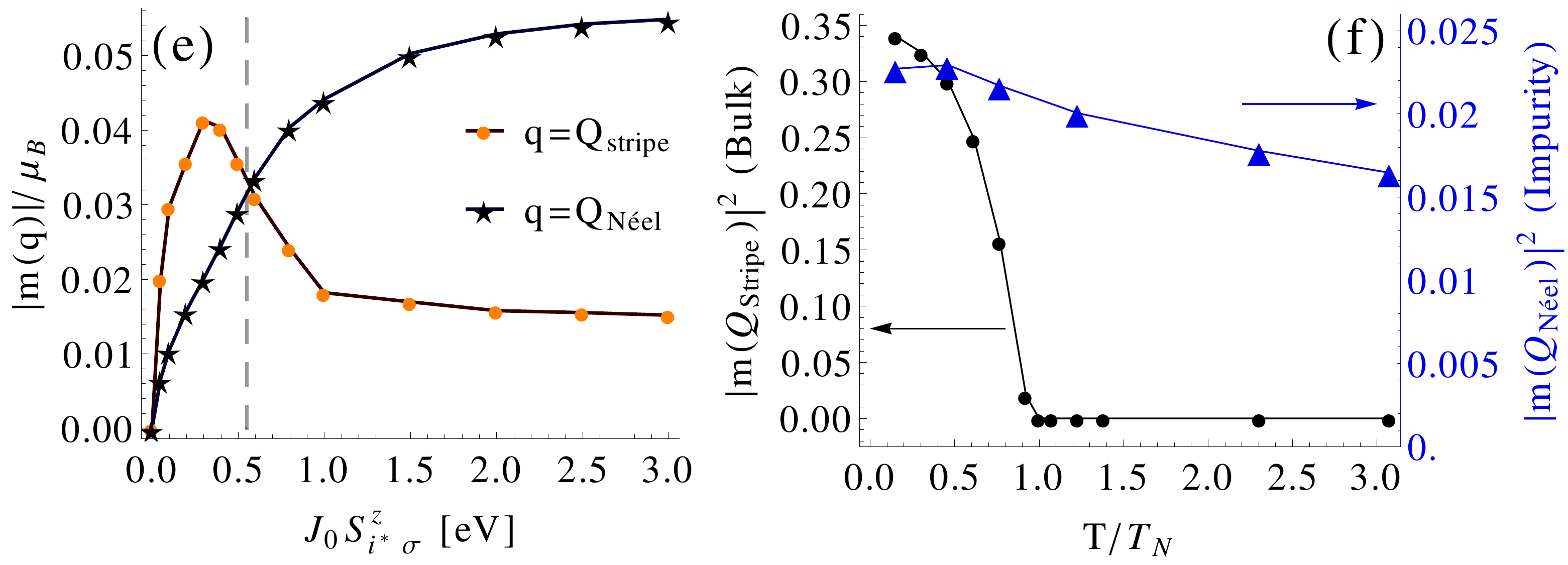}
\end{center}
\caption{(Color online) Real space magnetization $m(\mathbf{r})$ for (a) $J_0S^z_{\mathbf{i^*}\sigma}=0.3$ eV (weak) and (c) $J_0S^z_{\mathbf{i^*}\sigma}=0.8$ eV (strong) impurities. 
(b) and (d) $|m(\mathbf{q})|/\mu_B$ signal for the impurities in (a) and (b), respectively. 
(e) $|m(\mathbf{q})|/\mu_B$ at $\mathbf{q}=\mathbf{Q}_{stripe}$ (orange dots) and $\mathbf{q}=\mathbf{Q}_{N\acute{e}el}$ (black stars) as a function of impurity strength. 
The gray dashed line divides the weak and strong impurity regimes (see text).
$T/T_N=1.2$ for all these panels. 
(f) Temperature dependence of the intensity $|m(\mathbf{q})|^2/\mu_B^2$ at $\mathbf{Q}_{stripe}$ for the homogeneous system (black dots) and at $\mathbf{Q}_{N\acute{e}el}$ for the strong impurity (blue triangles).}
\label{fig:1}
\end{figure}

We start the discussion with the single-impurity effects above the homogeneous ordering temperature $T_N$. The spin polarization of the surrounding electrons induced by a weak impurity is shown in Fig.~\ref{fig:1}(a). Its structure can be mainly described by $\mathbf{Q}_x=(\pi,0)$ and $\mathbf{Q}_y=(0,\pi)$ stripe-type order along the $x$ and $y$ directions, respectively. The amplitude of the spatial oscillations rapidly weaken and vanish at a scale of $\sim 5-7$ lattice sites. The Fourier transform of the induced magnetization $|m(\mathbf{q})|$ shown in Fig.~\ref{fig:1}(b) exhibits sharp peaks at the $\mathbf{Q}_{stripe}$ wavevectors, arising from the real-space stripes. This local response of weak impurities simply reflects the structure as the spin susceptibility of the clean system, as expected from linear response theory. Upon increasing the magnetic impurity potential, however, the local response shown in Fig.~\ref{fig:1}(c,d) exhibits mainly $\mathbf{Q}_{N\acute{e}el}$ structure in the vicinity of the defect, and a weakened stripe order in the farther tails of the induced spin polarization. The broad $\mathbf{Q}_{N\acute{e}el}$ peak indicates shorter-range ($\pi,\pi$) order compared to the $\mathbf{Q}_{stripe}$ peaks characteristic of weak impurities [Fig.~\ref{fig:1}(b)]. The full evolution of the local response is presented in Fig.~\ref{fig:1}(e) showing the intensity of both the $\mathbf{Q}_{stripe}$ and $\mathbf{Q}_{N\acute{e}el}$ wavevectors as a function of impurity strength, revealing the crossover from $\mathbf{Q}_{stripe}$ order to mainly $\mathbf{Q}_{N\acute{e}el}$ order in the strong impurity limit. Note, however, that even the very strong impurities exhibit some weight at $\mathbf{Q}_{stripe}$ from surviving weak stripe-like modulations in the tails of the polarization cloud. 
Finally, Fig.~\ref{fig:1}(f) shows the $T$ dependence of the induced polarization of a strong impurity at $\mathbf{Q}_{N\acute{e}el}$ compared to the $\mathbf{Q}_{stripe}$ amplitude of the homogeneous system. 
The impurity nucleates a checkerboard polarization which is weakly $T$-dependent and persists well beyond $T_N$ consistent with $\mu$SR,\cite{inosov13} NMR,\cite{texier12} and neutron scattering.\cite{tucker12}. 

\begin{figure}
\begin{center}
\includegraphics[width=8.4cm]{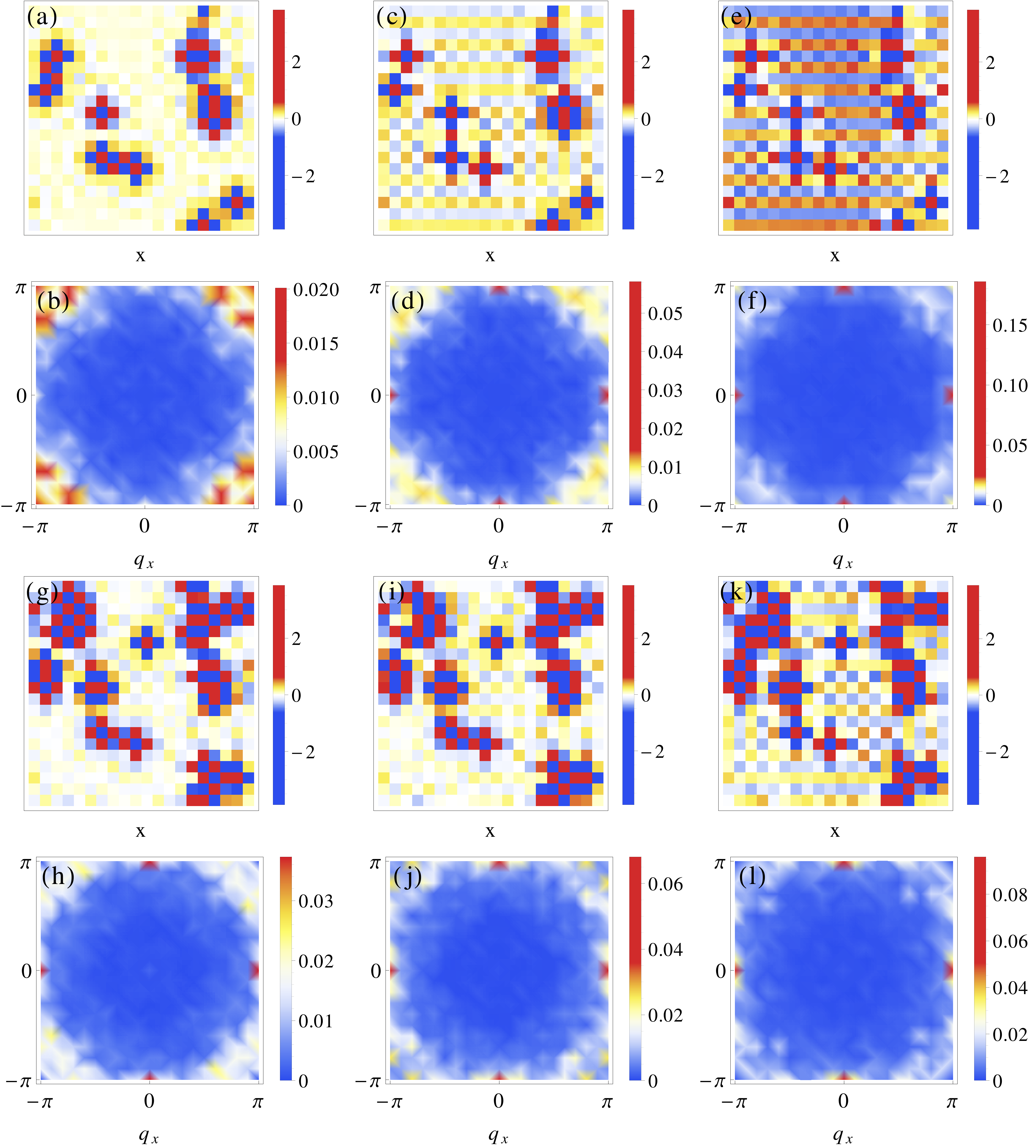}
\end{center}
\caption{(Color online) (a,c,e) Examples of $m(\mathbf{r})/\mu_B$ for a configuration with $x=3.0\%$ strong impurities ($J_0S^z_{\mathbf{i^*}\sigma}=\pm0.8$ eV) at (a) $T/T_N=1.08$, (c) 0.77 and (e) 0.46. 
(b,d,f) Corresponding $|m(\mathbf{q})|^2/\mu_B^2$ after averaging over eight distinct configurations with the same concentration. 
(g-l) The same as shown in (a-f) for a larger concentration of $x=6.0\%$. }
\label{fig:2}
\end{figure}

We now turn to the many-impurity case. The presence of both $\mathbf{Q}_{N\acute{e}el}$ and  $\mathbf{Q}_{stripe}$ magnetization induced by strong impurities makes them relevant from an experimental point of view\cite{tucker12,texier12,leboeuf13}. Below, we focus therefore on the $J_0 S^z_{\mathbf{i^*}\sigma}=\pm0.8$ eV defects [Fig.~\ref{fig:1}(c,d)]. Consider first the dilute case of $3\%$ disorder. Figure~\ref{fig:2}(a,c,e) show examples of the resulting real-space magnetization and Fig.~\ref{fig:2}(b,d,f) display the corresponding configuation-averaged $|m(\mathbf{q})|^2$. Above $T_N$ [Fig.~\ref{fig:2}(b)] the total signal in $q$-space is peaked around $\mathbf{Q}_{N\acute{e}el}$, similar to the single-impurity result for this kind of defect, whereas below $T_N$ [Fig.~\ref{fig:2}(d,f)] the broad $\mathbf{Q}_{N\acute{e}el}$ scattering coexists with the sharp $\mathbf{Q}_{stripe}$ order in agreement with experiments.\cite{tucker12} Now, consider a doubling of the impurity concentration as shown in Fig.~\ref{fig:2}(g-l). Below $T_N$ [Fig.~\ref{fig:2}(j,l)] the same dual nature is evident of sharp $\mathbf{Q}_{stripe}$ and broad $\mathbf{Q}_{N\acute{e}el}$ scattering as in the dilute case. Notably, however, above $T_N$ [Fig.~\ref{fig:2}(h)] the response is now dominated by sharp $\mathbf{Q}_{stripe}$ order as opposed to the result shown in Fig.~\ref{fig:2}(b). The cooperative impurity effect yielding this  long-range $\mathbf{Q}_{stripe}$ order is remarkable and not obvious from the corresponding real-space magnetization in Fig.~\ref{fig:2}(g). The origin of a critical Mn concentration $x_c$ needed for the emergence of high-$T$ $\mathbf{Q}_{stripe}$ order is evident within the present scenario; for the cooperative impurity effect to be relevant, the inter-impurity distance must be comparable to the size of the induced spin polarization cloud. The critical Mn concentration reported in Refs.~\onlinecite{kim10} and \onlinecite{inosov13} is $x_c\sim0.1$ implying that the magnetic $\mathbf{Q}_{stripe}$ tails induced by Mn in the real systems is slightly shorter ranged than for the parameters used in Fig.~\ref{fig:2}. 

\begin{figure}[b]
\begin{center}
\includegraphics[width=8.4cm]{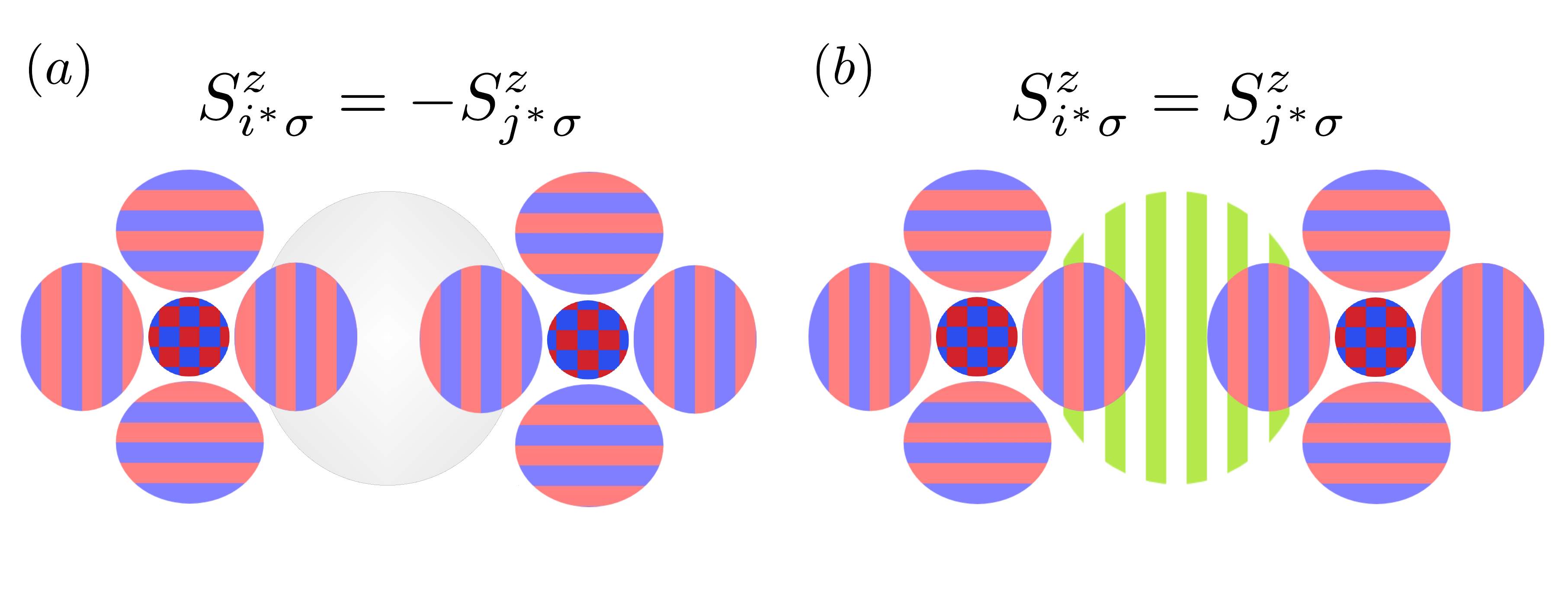}
\includegraphics[width=8.4cm]{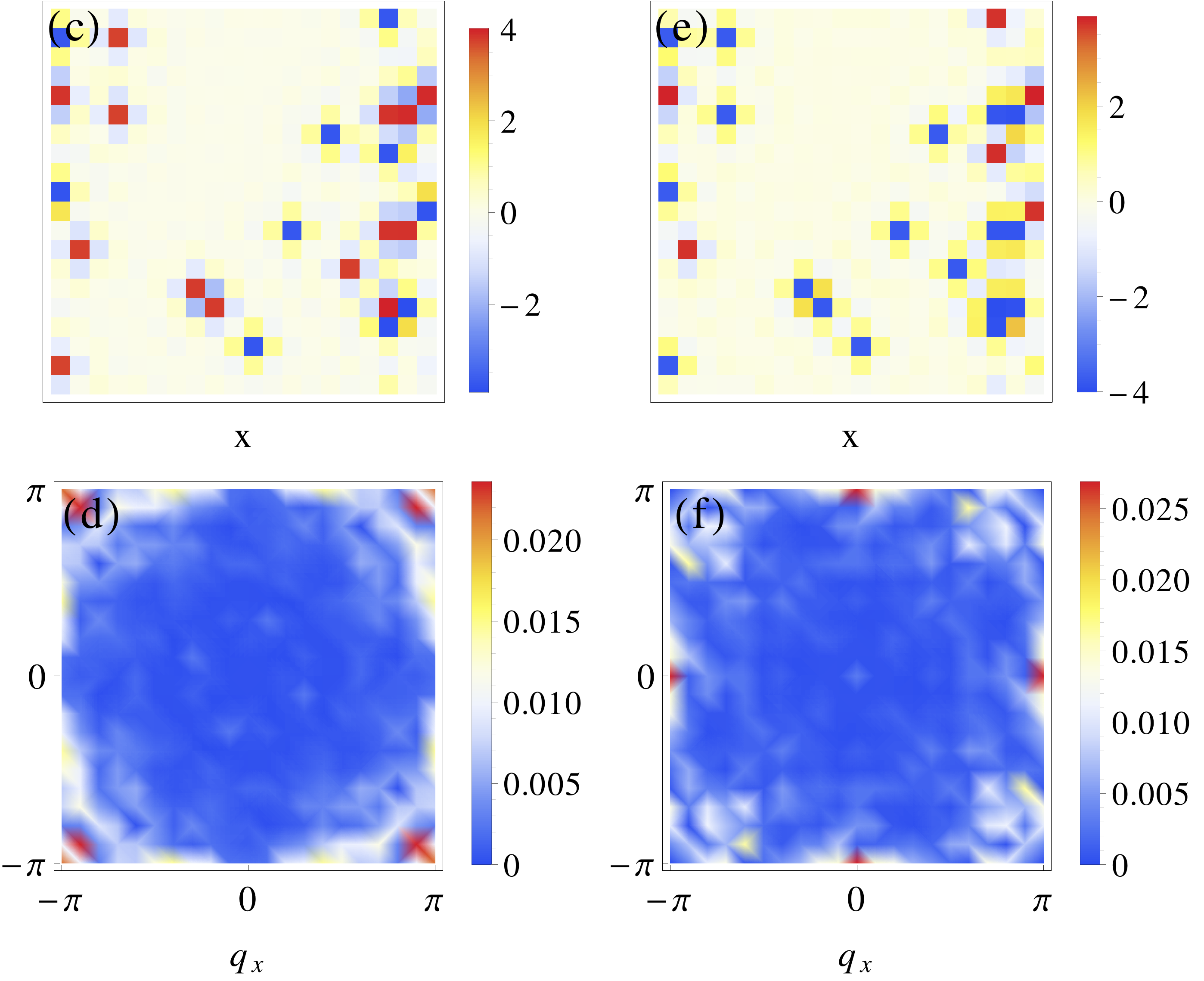}
\end{center}
\caption{(Color online) (a,b) Illustration of the order-by-disorder caused by constructive interference between two magnetic impurities; (a) destructive and (b) constructive interference. 
In both cases the impurities are separated by an even number of lattice sites. (c-f) Comparison of the magnetization in the case of "quenched" (a,b) impurity spins versus the "relaxed" case obtained by energy minimisation caused by allowed flipping of the impurity spins (c,d).}
\label{fig:3}
\end{figure}

To understand this cooperative behavior let us focus first on the simpler situation of two impurities at sites $\mathbf{i^*}$ and $\mathbf{j^*}$. Figure \ref{fig:3}(a,b) illustrate two cases where the spins are separated by an even number of lattice sites. 
In Fig.~\ref{fig:3}(b) $S^z_{\mathbf{i^*}\sigma}=S^z_{\mathbf{j^*}\sigma}$, there is constructive interference of their $\mathbf{Q}_{stripe}$ spin polarization, and magnetic stripes are induced in the surrounding conduction electrons. The opposite case of $S^z_{\mathbf{i^*}\sigma}=-S^z_{\mathbf{j^*}\sigma}$ with destructive interference is shown in Fig.~\ref{fig:3}(a). Importantly, the configuration with the lowest free energy $\mathcal{F}$ is the one with \emph{constructive} interference of the spin-polarized electrons capable of generating inter-impurity regions of $\mathbf{Q}_{stripe}$ order. In the corresponding many-impurity case, the importance of optimizing the impurity spin orientations is shown in Figs.~\ref{fig:3}(c-f). Here we compare the case of randomly chosen spin orientations (quenched case) [Figs.~\ref{fig:3}(c,d)] with the lower energy annealed case where the impurity spins are allowed to orient themselves favorably to the spin polarization of their neighbors [Figs.~\ref{fig:3}(e,f)]. Evidently, the induced  $\mathbf{Q}_{stripe}$ order exists only in the latter situation. 

The results shown in Figs.~\ref{fig:2}-\ref{fig:3} constitute an example of order-from-disorder. It is qualitatively similar to the interactions between impurities in quantum spin chains which are non-frustrating because of the freedom of impurity spins to reorient themselves according to the neighboring AF induced impurity clouds and thereby lower the exchange energy.\cite{shender91} A similar scenario has also been proposed for the spin-glass phase of the cuprates, where non-magnetic dopant-induced AF clouds overlap and form a network of quasi-long-range $(\pi,\pi)$ order.\cite{andersen07} 

\begin{figure}[t]
\begin{center}
\includegraphics[width=7.2cm]{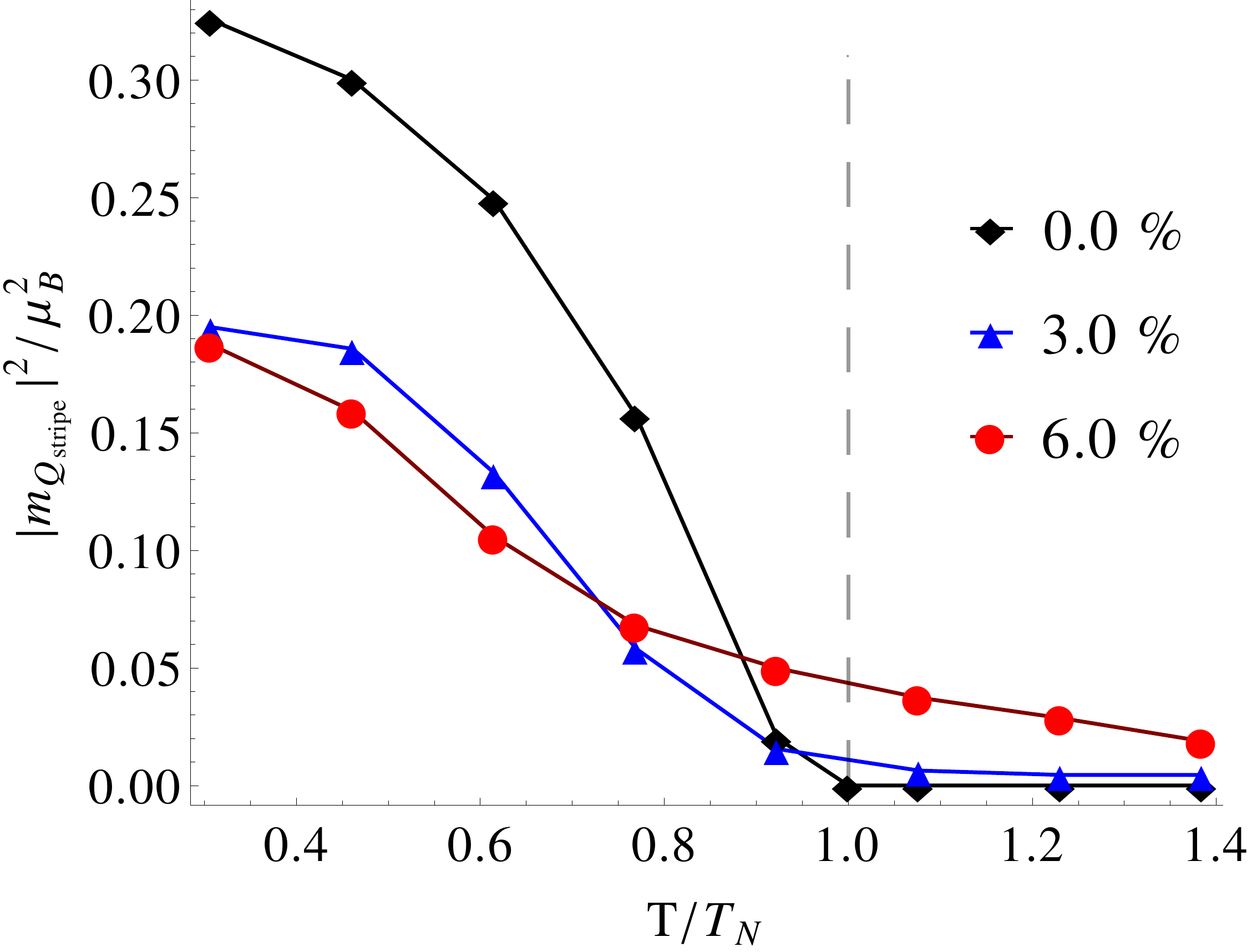}
\end{center}
\caption{(Color online) Temperature dependence of the magnetic $\mathbf{q}=\mathbf{Q}_{stripe}$ Bragg peak for the homogeneous system (black diamonds), $x=3.0\%$ (blue triangles), and $x=6.0\%$ (red dots) disorder. Vertical dashed line indicates $T=T_N$.}
\label{fig:4}
\end{figure}

We end the discussion with the $T$ dependence of the $|m(\mathbf{Q}_{stripe})|^2$ magnetic signal. Figure~\ref{fig:4} shows the intensity of the peak for $3\%$ and $6\%$ disorder. In the $x<x_c$ case, the signal is lost at $T\lesssim T_N$ and it exhibits a clear suppression of weight at low $T$ compared to the clean system. From Figs.~\ref{fig:2}(d,f) it is evident that most of this weight has been transferred to the $\mathbf{Q}_{N\acute{e}el}$ wavevector. The $x>x_c$ case shows a smeared transition as opposed to the sharp transition of the clean system.
The disorder induced $\mathbf{Q}_{stripe}$ signal persists to $T$ significantly above the ordering temperature of the clean system, $T/T_N\sim1.4$, where the signal intensity has been suppressed by $\sim90\%$  with respect to its low-$T$ value consistent with elastic neutron measurements.\cite{kim10,inosov13} 

Finally we comment on the reported absence of orthorhombic distortion in Mn-doped Ba-122.\cite{kim10} Of course, since we do not explicitly include a coupling to the lattice within the model, we cannot make quantitative statements. However, assuming that the orthorhombic distortion of the clean Mn-free system is caused by a magneto-elastic coupling, it is evident from e.g. Figs.~\ref{fig:2}(g-k) that since sizeable domains of neither $(\pi,0)$ nor $(0,\pi)$ order exist even at the lowest $T$ when the Mn doping approaches $x_c$, the same magneto-elastic coupling should cause only a short-scale mixture of intertwined tetragonal and orthorhombic regions consistent with Ref.~\onlinecite{inosov13}.

In summary, we have studied the cooperative effects of magnetic impurities within a realistic five-band model relevant for the iron pnictides. The resulting induced long-range magnetic stripe order of the conduction electrons constitute an example of an order-from-disorder phenomenon, which explains the main experimental observations of the magnetic properties of Mn doped BaFe$_2$As$_2$. This includes the presence of a high-$T$ $\mathbf{Q}_{stripe}$ stripe magnetic phase beyond a critical Mn concentration, the presence of short-range checkerboard $(\pi,\pi)$ spin fluctuations, and the absence of a clear tetragonal to orthorhombic structural transition.
 
We thank P. J. Hirschfeld,  F. Kr\"{u}ger, and J. Paaske for useful discussions. We acknowledge support from a Lundbeckfond fellowship (Grant A9318).

\end{document}